\documentclass[12pt,letterpaper]{article}
\usepackage{epsfig,rotating,setspace,latexsym,amsmath,epsf,amssymb,bm}
\usepackage{cite,graphicx,authblk,color}

\title{Minimum CSIT  to achieve Maximum Degrees of Freedom for the MISO BC\thanks{E-mail: tandonr@vt.edu, syed@uci.edu, sshlomo@ee.technion.ac.il}}
\vspace{3cm}
\author[1]{Ravi Tandon}
\author[2]{Syed Ali Jafar}
\author[3]{Shlomo Shamai}
\affil[1]{\small Department of ECE, Virginia Tech, Blacksburg, VA, USA.}
\affil[2]{\small Department of EECS, University of California, Irvine, CA, USA.}
\affil[3]{\small Department of EE, Technion, Israel Institute of Technology, Haifa, Israel.}

\newtheorem{Theo}{Theorem}
\newtheorem{remark}{Remark}
\newtheorem{Lem}{Lemma}

\begin{document}
\maketitle
\thispagestyle{empty}
\vspace{-0.5cm}
\begin{abstract}
Channel state information at the transmitter (CSIT) is a key ingredient in realizing 
the multiplexing gain provided by distributed MIMO systems. For a downlink  
multiple-input single output (MISO) broadcast channel, with $M$ antennas at the transmitters
and $K$ single antenna receivers, the maximum multiplexing gain or the maximum degrees of freedom (DoF) is $\min(M,K)$. 
The optimal DoF of $\min(M,K)$ is achievable if the transmitter has access to perfect, instantaneous CSIT from all receivers. 
In this paper, we pose the question that what is {\emph{minimum amount}} of CSIT required per user in order to achieve the 
maximum DoF of $\min(M,K)$. By minimum amount of CSIT per user, we refer to the \textit{minimum fraction of time} that the transmitter 
has access to perfect and instantaneous CSIT from a user. Through a novel converse proof and an achievable scheme, it is shown
that the minimum fraction of time, perfect CSIT is required per user in order to achieve the DoF of $\min(M,K)$ is given by $\min(M,K)/K$. 
\end{abstract}
\section{Introduction}
The popularity of multiple-input multiple-output (MIMO) systems in current wireless systems 
stems from the fact that MIMO exploits the space dimension to provide capacity gains and reliability.  Among one of the gains provided by the use of MIMO is for the case of downlink transmission to several distributed receivers, each desiring independent information from the base station. For a $M$-antenna transmitter, and $K$ single antenna receivers, it is well known that the maximum pre-log of capacity (often referred to as the degrees of freedom (DoF)) is $\min(M,K)$. This implies that the use of multiple antennas at the transmitters exploits the channel diversity and compensates for the fact that the receivers are distributed in space. However, creating such $\min(M,K)$ non-interfering data \textit{streams} to the receivers requires timely and accurate channel state information at the transmitter (CSIT) in order to employ pre-coding techniques at the transmitter. This requirement necessitates timely channel estimation at the receivers and reliable feedback channels to the transmitters and may result in significant overhead in the uplink. 

It is therefore of significant practical importance to understand and quantify the perfect CSIT requirements in order to reap the gains provided by MIMO systems. To this end, we focus on a simplistic yet practical model: the MISO broadcast channel with $M$ transmit antennas and $K$ single antenna receivers. We address and answer the following question: what is the minimum 
amount of perfect CSIT (per receiver) required in order to achieve the maximum DoF of $\min(M,K)$. Formally, we model the problem as follows: transmitter can choose to acquire perfect CSIT from a specific receiver for a fraction $\lambda$ of the total communication duration.
Furthermore, the instances during which the transmitter receives perfect CSIT from two distinct receivers may or may not be the same; and we permit this flexibility as a design parameter. In the remaining $(1-\lambda)$ fraction of time, we assume that the transmitter may either have  completely delayed CSIT or it may have no CSIT. The symmetric assumption of having the same fraction $\lambda$ for all receivers is well motivated from two aspects: a) under a total cost constraint of acquiring CSIT, the best allocation would correspond to allocating equal resources in the uplink, and b) this assumption would automatically ensure fairness in the achievable DoF per-user (which is a consequence of one of our results).  

The main result of this paper is to show that $\lambda^{*}(M,K)= \frac{\min(M,K)}{K}$, for $\min(M,K)>1$. The main technical contribution is the converse where we present an outer bound to the DoF region for the model described above.  The outer bound presented in the paper holds valid for a stronger setting in which the remaining $(1-\lambda^{*})$ fraction of time, the transmitter always has delayed CSIT.  On the other hand, the achievability proof of using $\min(M,K)/K$ fraction of CSIT to achieve $\min(M,K)$ DoF requires no CSIT from the remaining $(1-\lambda^{*})$ fraction of the time from each of the receiver. 

Related work: There is significant amount of literature investigating various forms of channel state information at the transmitters in the context of MIMO broadcast channels: including full (perfect and instantaneous) CSIT \cite{MIMOBC}, no CSIT \cite{CaireShamai,NoCSITJafar,VV:NOCSIT,JafarGoldsmithIsotropic},  delayed CSIT \cite{MaddahAli-Tse:DCSI-BC,VV:DCSI-BC}, compound CSIT \cite{Weingarten_Shamai_Kramer,CompoundJafar,MAMA:CompoundBC}, quantized CSIT \cite{Jindal_BCFB, Caire_Jindal_Shamai,Kobayashi_Caire_Jindal},  mixed (perfect delayed and partial instantaneous) CSIT \cite{JafarTCBC,KobayashiTCBC,EliaMixed}, asymmetric CSIT (perfect CSIT for one user, delayed CSIT for the other) \cite{Jafar_corr, RetroIA} and with knowledge of only the channel coherence patterns available to the transmitter \cite{Jafar_corr, Wang_Gou_Jafar}. 

This work is  related to a new model of alternating CSIT, which was proposed in \cite{TandonJafarShamaiPoor:alternating} in the context of MISO broadcast channel with two receivers. As a consequence of the key result presented in \cite{TandonJafarShamaiPoor:alternating}, it can be shown that for the case of $M=K=2$, the minimum amount of perfect CSIT (per-user) required to achieve the maximum DoF of $2$ is $1$. 
In a recent work of Lee and Heath  \cite{NamyoonHeath}, the $K$ receiver MISO BC with $(K-1)$ transmit antennas is considered and a novel achievable scheme is proposed that utilizes perfect CSIT for $\frac{(K-1)}{K}$ fraction of time and delayed CSIT for the remaining $\frac{1}{K}$ fraction of time and achieves the maximum DoF of $(K-1)$. Our converse result can be regarded as a complement to the result of \cite{NamyoonHeath} by showing that $(K-1)/K$ is indeed the \textit{minimum} amount of fraction of perfect CSIT required in order to achieve $(K-1)$ DoF.   

\section{System Model}
A multiple-input single-output (MISO) broadcast channel is considered in which a transmitter (with $M$ transmit antennas) wishes to send $K$ independent messages $W_{1},\ldots,W_{K}$ to  $K$ receivers, where the message $W_{k}$ is intended for the $k$th receiver, and each receiver is equipped with a single antenna. The channel input output relationships are given as:
\begin{align}
Y_{k}(t)= \mathbf{H}_{k}(t)\mathbf{X}(t) + N_{k}(t), \quad k=1,\ldots,K,
\end{align}
where $Y_{k}(t)$ is the scalar channel output of receiver $k$ at time $t$,
$\mathbf{X}(t)$ is the $M\times 1$ channel input at time $t$ which satisfies the power constraint $E[||\mathbf{X}(t)||^{2}]\leq P$, $N_{k}(t)\sim \mathcal{CN}(0,1)$ is a 
circularly symmetric complex additive white Gaussian noise at receiver $k$ at time $t$. The $M\times 1$ channel vectors $\mathbf{H}_{k}(t)$ to receiver $k$ are independent and
identically distributed (i.i.d.) with continuous distributions, and are also i.i.d. over time. 
We assume that the receivers have global channel state information (CSIR). 

From each receiver $k$, the transmitter can have access to either perfect, delayed or no CSIT.
We denote this CSIT availability for each receiver $k$ at time $t$ as:
\begin{align}
\mu_{k}(t)=
\begin{cases}
P, &\text{ perfect CSIT}, \\
D, &\text{ delayed CSIT}, \\
N, &\text{ no CSIT}.
\end{cases}
\end{align}
In this paper, we consider the case in which the fraction of time the transmitter has perfect CSIT
from the $k$th receiver is at most $\lambda$, for all $k=1,\ldots,K$, i.e., for a total communication period of $n$ channel uses, we must have
\begin{align}
\frac{\sum_{t=1}^{n} \mathbb{I}({\mu_{k}(t)=P})}{n}\leq \lambda.
\end{align}

The rate tuple $(R_{1},\ldots,R_{K})$, with $R_{k}=\log(|W_{k}|)/n$, where $n$ is the number of channel uses, is achievable if the probability of decoding error for $k=1\ldots,K$ can be made
arbitrarily small for sufficiently large $n$.  The degrees of freedom region $\mathcal{D}(\lambda)$, is defined as the closure of the set of all achievable tuples $(d_{1},\ldots,d_{K})$, with $d_{k}=\lim_{P\rightarrow \infty} \frac{R_{k}}{\log(P)}$. 

Furthermore, we denote the maximum sum DoF as:
\begin{align}
DoF^{*}(\lambda)&= \max_{\{d_{i}\}\in \mathcal{D}(\lambda)} d_{1}+\ldots+d_{K}
\end{align} 
It is clear that $DoF^{*}(\lambda)$ is a non-decreasing function of $\lambda$ and it is upper bounded by $\min(M,K)$. In the next section, we characterize the minimum value of $\lambda$
for which $DoF^{*}(\lambda)=\min(M,K)$. 

\section{Main Results}
\begin{Theo}\label{Theorem1}
The minimum fraction of perfect CSIT per-user to achieve the maximum DoF of $\min(M,K)$ for the $K$ user MISO broadcast channel 
is given by 
\begin{align}
\lambda^{*}(M,K)=
\begin{cases}
0, & \min(M,K)=1\nonumber\\
\frac{\min(M,K)}{K}, &\min(M,K)>1.\nonumber
\end{cases}
\end{align}
\end{Theo}

Note that Theorem \ref{Theorem1} is trivial for $\min(M,K)=1$ since with full CSIR and no CSIT, $1$ DoF is achievable, and thus $\lambda^{*}=0$. The interesting case is when $\min(M,K)>1$ and we prove Theorem \ref{Theorem1} for this case in two parts:
\begin{itemize}
\item We present a simple achievable scheme which utilizes a fraction of $\lambda=\frac{\min(M,K)}{K}$ amount of CSIT
per user and achieves the maximum DoF of $\min(M,K)$. This would show that $\lambda^{*}(M,N)\leq \frac{\min(M,K)}{K}$.
\item We present an outer bound to the DoF region of the $K$ user MISO broadcast channel; which is a function of $\lambda$. 
From this outer bound, we then show that $\lambda^{*}(M,N)\geq \frac{\min(M,K)}{K}$. The main contribution is the proof of the lower bound. 
\end{itemize}

\subsection{Achieving $\min(M,K)$ DoF with $\lambda = \min(M,K)/K$}
We first illustrate the proof through an example. Consider the case in which $M=2$, $K=3$ so that $\min(M,K)=2$.
We want to use perfect CSIT from each user for a $2/3$-fraction of the total communication period. 
Consider the following block scheme (of block length $3$) for any $i\geq 1$:
\begin{align}
t=i: &\text{ Perfect CSIT from receivers } 1,2, \text{ No CSIT from receiver } 3.\nonumber\\
t=i+1: &\text{ Perfect CSIT from receivers } 2,3, \text{ No CSIT from receiver } 1.\nonumber\\
t=i+2: &\text{ Perfect CSIT from receivers } 1,3, \text{ No CSIT from receiver } 2.\nonumber
\end{align}
Clearly, at each time $t$, sum DoF of $2$ is achievable by using $M=2$ transmit antennas and having perfect CSIT from two distinct receivers. 
Thus, this scheme achieves a DoF of $2$.  In any given block, the number of instances transmitter obtains perfect CSIT from a receiver is $2$, and the length of the block is $3$. The fraction of time for which perfect CSIT is required from the $k$th receiver is $2/3$.
\begin{figure}[t]
  \centering
\includegraphics[width=11.5cm]{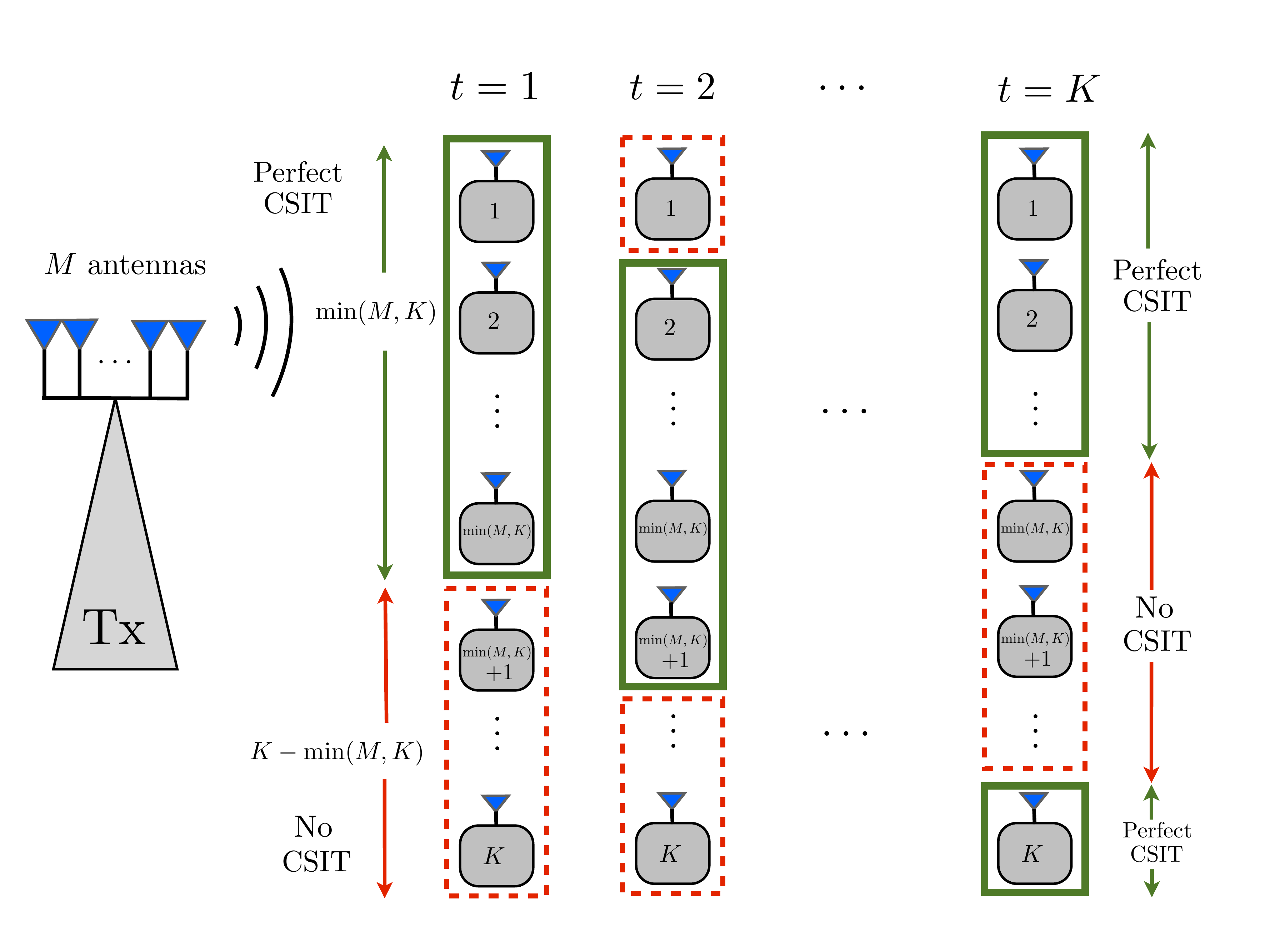}
\caption{Achieving $\min(M,K)$ DoF with $\lambda= \min(M,K)/K$.}\label{Figure1}
\vspace{-0.1in}
\end{figure}
We note that the scheme proposed above that requires perfect CSIT $\lambda=2/3$ fraction of time is not necessarily unique.
Another scheme that achieves $2$ DoF has been proposed by Lee and Heath requires the following CSIT pattern\footnote{We note here that \cite{NamyoonHeath} assumes a block fading model in which the channel to the receivers remains constant for a block of $T_{c}$ time slots; and the channel is known to the transmitter after $T_{f}\leq T_{c}$ time slots. The key idea is to use the channel instances across different blocks (channels across blocks are i.i.d.) and achieve the min-cut value of $2$ DoF for $T_{c}=3$ and $T_{f}=1$, which is equivalent to $\lambda=2/3$ for our system model. } \cite{NamyoonHeath}:
\begin{align}
t=i: &\text{ Delayed CSIT from receivers 1, 2, 3}.\nonumber\\
t=i+1: &\text{ Perfect CSIT from receivers 1, 2, 3}.\nonumber\\
t=i+2: &\text{ Perfect CSIT from receivers 1, 2, 3}.\nonumber
\end{align} 
We next present the proof for $\lambda^{*}(M,K)\leq \min(M,K)/K$ for arbitrary $M$ and $K$. 
We consider a scheme of block length $K$ with the following CSIT pattern:
\begin{align}
t=1&: \text{ Perfect CSIT from receivers } 1, 2, 3,\ldots, \min(M,K).\nonumber\\
t=2&: \text{ Perfect CSIT from receivers } 2, 3, 4,\ldots, \min(M,K)+1.\nonumber\\
t=3&: \text{ Perfect CSIT from receivers } 3, 4, 5,\ldots, \min(M,K)+2.\nonumber\\
&\hspace{-0.5cm}\vdots \nonumber\\
t=K&: \text{ Perfect CSIT from receivers } K,1,2,\ldots, \min(M,K)-1\nonumber.
\end{align}
At each time instant, perfect CSIT is present from $\min(M,K)$ receivers and no CSIT from the remaining $K-\min(M,K)$ receivers. A sum DoF of $\min(M,K)$ is achievable
at each time instant and therefore a sum DoF of $\min(M,K)$ is also achievable for this scheme. 
The fraction of time perfect CSIT is obtained from any specific receiver is $\min(M,K)/K$ and therefore we have shown that 
$\lambda^{*}(M,K)\leq \min(M,K)/K$. 

Figure \ref{Figure1} shows a useful way to interpret this scheme. Consider a \emph{window} of $\min(M,K)$ users. The transmitter only requests perfect CSIT from the users falling in this window; and then cyclically shifts this window $K$ times. Each user falls in the window a total number of $\min(M,K)$ times; so that the fraction of CSIT required per user is $\min(M,K)/K$. 
Note that for $\min(M,K)=K$, there is only one such window spanning all the users and thus $\lambda^{*}=1$, when $\min(M,K)=K$. 

\subsection{Converse Proof}
We present an outer bound to the DoF region of the $K$-user MISO BC in which perfect CSIT
is available from the $k$th user for $\lambda$ fraction of time. For the remaining $(1-\lambda)$ fraction of time, 
the transmitter could have access to either delayed CSIT or no CSIT from the $k$th user. Furthermore, we make no assumptions 
about how the instances during which the transmitter has access to perfect/delayed/no CSIT from a specific user relate to the 
instances it has access to perfect/delayed/no CSIT from the remaining users. That is, the {\emph{only}} assumption made is that from each 
user, the fraction of time perfect CSIT is available is $\lambda$; and such instances could be arbitrarily distributed 
across the communication period. 
\begin{Lem}\label{Lemma1}
An outer bound for the $K$-user MISO BC with perfect CSIT from each user for $\lambda$-fraction of time is given as:
\begin{align}
Md_{1}+ d_{2}+\ldots + d_{K}&\leq M+ (\min(M,K)-1)\lambda\\
d_{1}+ Md_{2}+\ldots + d_{K}&\leq M+ (\min(M,K)-1)\lambda\\
&\vdots\\
d_{1}+ d_{2}+\ldots + Md_{K}&\leq M+ (\min(M,K)-1)\lambda.
\end{align}
\end{Lem}
We prove Lemma \ref{Lemma1} in the Appendix. 

Summing up all the bounds in Lemma \ref{Lemma1}, we obtain:
\begin{align}
d_{1}+d_{2}+\ldots+d_{K}&\leq \frac{K\big[M+(\min(M,K)-1)\lambda\big]}{M+K-1}
\end{align}
We are interested in the case when $d_{1}+d_{2}+\ldots+d_{K}=\min(M,K)$, and setting DoF$=\min(M,K)$ in the l.h.s. above, we get
\begin{align}
\min(M,K)&\leq  \frac{K\big[M+(\min(M,K)-1)\lambda\big]}{M+K-1}.\label{Mainbound}
\end{align}

We now consider two cases: 
\begin{itemize}

\item Case I: $\min(M,K)=K$

In this case, (\ref{Mainbound}) simplifies to
\begin{align}
K&\leq \frac{K(M+(K-1)\lambda)}{M+K-1},
\end{align}
which leads to 
\begin{align}
M+(K-1) &\leq M+(K-1)\lambda
\end{align}
which gives the bound
\begin{align}
\lambda&\geq 1.
\end{align}

\item Case II: $\min(M,K)=M$

In this case, (\ref{Mainbound}) simplifies to
\begin{align}
M&\leq \frac{K(M+(M-1)\lambda)}{M+K-1},
\end{align}
which leads to 
\begin{align}
KM+M(M-1) &\leq KM+K(M-1)\lambda,
\end{align}
which gives the bound
\begin{align}
\lambda&\geq \frac{M}{K}.
\end{align}
\end{itemize}
Thus, from Cases I and II, we have the following lower bound:
\begin{align}
\lambda^*(M,K)&\geq \frac{\min(M,K)}{K}.
\end{align}
This completes the proof of Theorem \ref{Theorem1}.

\section{Conclusions}
In this paper, the notion of minimum fraction of time for which perfect CSIT is needed in order to
achieve the maximum DoF is captured and investigated in the context of mutli-receiver MISO broadcast channels. A novel converse proof is presented which gives an outer bound to the DoF region of the MISO BC as a function of $\lambda$, the fraction of time perfect CSIT is present from each receiver. We note that the outer bounds let us correctly characterize the minimum CSIT to achieve maximum DoF; however the question of optimality of the outer bound for values of $\lambda$ other than $\lambda^{*}$ is an interesting one. 

\section{Appendix: Proof of Lemma \ref{Lemma1}}
Since all the bounds are symmetric, it suffices to prove one of them. Therefore, we present the proof of the following bound:
\begin{align}
Md_{1}+ d_{2}+\ldots + d_{K}&\leq M+ (\min(M,K)-1)\lambda
\end{align}
To prove this bound, we start by considering the original MISO BC.
Consider the channel output $Y_{1}^{n}$ at receiver $1$ and we further denote $Y_{1}^{n}$ as follows:
\begin{align}
Y_{1}^{n}&= \left(Y_{1,P}^{n}, Y_{1,NP}^{n}\right),
\end{align}
where 
\begin{itemize}
\item $Y_{1,P}^{n}$: channel outputs at receiver $1$ for instances in which perfect CSIT is {\emph{present}} from receiver $1$.
\item $Y_{1,NP}^{n}$: channel outputs at receiver $1$ for instances in which perfect CSIT is {\emph{not present}} from receiver $1$.
\end{itemize}
We next arrange the channel outputs of the remaining $(K-1)$ receivers and denote these as:
\begin{align}
Y_{k}^{n}&= \left(Y_{k,(1,P)}^{n}, Y_{k,(1,NP)}^{n}\right), \quad k=2,\ldots, K,
\end{align}
where 
\begin{itemize}
\item $Y_{k, (1,P)}^{n}$: channel outputs at receiver $k$ for those instances in which perfect CSIT is {\emph{present}} from receiver $1$.
\item $Y_{k,(1,NP)}^{n}$: channel outputs at receiver $k$ for those instances in which perfect CSIT is {\emph{not present}} from receiver $1$.
\end{itemize}
We next enhance the original BC by colluding the outputs of receivers $2, \ldots, K$. 
That is, we now have two receivers: receiver $1$ and receiver $\tilde{2}$ with the following channel outputs:
\begin{itemize}
\item At receiver $1$: $\left(Y_{1,P}^{n}, Y_{1,NP}^{n}\right)$
\item At receiver $\tilde{2}$: $\left(Y_{2,(1,P)}^{n},\ldots, Y_{K,(1,P)}^{n}, Y_{2,(1,NP)}^{n},\ldots, Y_{K,(1,NP)}^{n}\right)$
\end{itemize}
We next further enhance this BC by giving the channel output of receiver $1$ to receiver $\tilde{2}$:
\begin{itemize}
\item At receiver $1$: $\left(Y_{1,P}^{n}, Y_{1,NP}^{n}\right)$
\item At receiver $\tilde{2}$: $\left(Y_{1,P}^{n}, Y_{2,(1,P)}^{n},\ldots, Y_{K,(1,P)}^{n}, Y_{1,NP}^{n},Y_{2,(1,NP)}^{n},\ldots, Y_{K,(1,NP)}^{n}\right)$
\end{itemize}
This is a physically degraded broadcast channel, for which it is known from \cite{ElGamalFB} that feedback does not increase the capacity region
(and hence the DoF region). Therefore, from the instances corresponding to $Y_{1,NP}^{n}$, we can remove the assumption of delayed CSIT
(if any) without effecting the capacity (and the DoF) region. 

We next introduce $(M-1)$ artificial receivers that are statistically indistinguishable from receiver $1$. 
We denote the outputs receivers at receiver $j$ as follows:
\begin{align}
\tilde{Y}_{j}^{n}= \left(Y_{1,P}^{n}, \tilde{Y}_{j,NP}^{n}\right), \quad j=1,\ldots, M-1,
\end{align}
where $Y_{1,P}^{n}$ is exactly the same as the channel output at receiver $1$ corresponding to instances with perfect CSIT; and the channel corresponding to $\tilde{Y}_{j,NP}^{n}$ is identically distributed as the channel output to receiver $1$ as in $Y_{1,NP}^{n}$. 

Let $\Omega$ denote the total channel state information of the original BC and that of the artificial receivers. 
We next note that in this enhanced physically degraded broadcast channel, the capacity region depends only on the marginals and thus 
if receiver $1$ can decode the message $W_{1}$; then all the artificial $M-1$ receivers must also be able to decode the message $W_{1}$. 

We start with the following sequence of bounds for receiver $1$:
\begin{align}
nR_{1}&= H(W_{1})\\
&= H(W_{1}|\Omega)\\
&\leq I(W_{1};Y_{1}^{n}|\Omega)+n\epsilon_{n}\\
&= I(W_{1};Y_{1,P}^{n},Y_{1,NP}^{n}|\Omega)+n\epsilon_{n}\\
&= h(Y_{1,P}^{n},Y_{1,NP}^{n}|\Omega) - h(Y_{1,P}^{n},Y_{1,NP}^{n}|W_{1}, \Omega) + n\epsilon_{n}\\
&\leq n\log(P) - h(Y_{1,P}^{n},Y_{1,NP}^{n}|W_{1}, \Omega) + n\epsilon_{n}\\
&= n\log(P) - h(Y_{1,P}^{n}|W_{1},\Omega) - h(Y_{1,NP}^{n}|W_{1}, Y_{1,P}^{n},\Omega) + n\epsilon_{n}\label{Eq:R1}
\end{align}
Similarly, for each of the artificial receiver $j$, we also have:
\begin{align}
nR_{1}&\leq n\log(P) - \underbrace{h(Y_{1,P}^{n}|W_{1},\Omega)}_{\geq no(\log(P))} - h(\tilde{Y}_{j,NP}^{n}|W_{1}, Y_{1,P}^{n},\Omega) + n\epsilon_{n}\\
&\leq n\log(P) - h(\tilde{Y}_{j,NP}^{n}|W_{1}, Y_{1,P}^{n},\Omega) + n\epsilon_{n} -no(\log(P))\label{Eq:Rj}
\end{align}
Adding these total $M$ bounds, we obtain
\begin{align}
nMR_{1}&\leq nM\log(P) -  h(Y_{1,P}^{n}|W_{1},\Omega) \nonumber\\
&\quad-  h(Y_{1,NP}^{n}|W_{1}, Y_{1,P}^{n},\Omega) -\sum_{j=1}^{M-1}h(\tilde{Y}_{j,NP}^{n}|W_{1}, Y_{1,P}^{n},\Omega) \nonumber\\
&\quad + nM\epsilon_{n}- no(\log(P))\\
&\leq nM\log(P) -  h(Y_{1,P}^{n}|W_{1},\Omega) \nonumber\\
&\quad-  h(Y_{1,NP}^{n},\tilde{Y}_{1,NP}^{n},\ldots, \tilde{Y}_{M-1,NP}^{n}|W_{1}, Y_{1,P}^{n},\Omega) \nonumber\\
&\quad + nM\epsilon_{n}- no(\log(P))\label{Eq:R1bound}
\end{align}

We next have the following sequence of bounds for the remaining receivers $2,\ldots, K$:
\begin{align}
&n(R_{2}+R_{3}+\cdots+R_{K} )\nonumber\\
&= H(W_{2},W_{3},\ldots,W_{K})\\
&= H(W_{2},W_{3},\ldots,W_{K}|W_{1},\Omega)\\
&\leq I(W_{2},W_{3},\ldots,W_{K};  Y_{2}^{n},\ldots,Y_{K}^{n}, Y_{1}^{n}, \tilde{Y}_{1}^{n},\ldots, \tilde{Y}_{M-1}^{n} |W_{1},\Omega) + n\epsilon_{n}\\
&\leq h(Y_{2}^{n},\ldots,Y_{K}^{n}, Y_{1}^{n}, \tilde{Y}_{1}^{n},\ldots, \tilde{Y}_{M-1}^{n} |W_{1},\Omega) + n\epsilon_{n}\\
&= h(Y_{1}^{n},\tilde{Y}_{1}^{n},\ldots, \tilde{Y}_{M-1}^{n} |W_{1},\Omega) +n\epsilon_{n}\nonumber\\
&\quad + h(Y_{2}^{n},\ldots,Y_{K}^{n}| Y_{1}^{n},\tilde{Y}_{1}^{n},\ldots, \tilde{Y}_{M-1}^{n} |W_{1},\Omega)\\
&= h(Y_{1,P}^{n}, Y_{1,NP}^{n},\tilde{Y}_{1,NP}^{n},\ldots, \tilde{Y}_{M-1,NP}^{n} |W_{1},\Omega) +n\epsilon_{n}\nonumber\\
&\quad + h(Y_{2}^{n},\ldots,Y_{K}^{n}| Y_{1}^{n},\tilde{Y}_{1}^{n},\ldots, \tilde{Y}_{M-1}^{n} ,W_{1},\Omega)\\
&\leq h(Y_{1,P}^{n}, Y_{1,NP}^{n},\tilde{Y}_{1,NP}^{n},\ldots, \tilde{Y}_{M-1,NP}^{n} |W_{1},\Omega) +n\epsilon_{n}\nonumber\\
&\quad + h\left(Y_{2,(1,P)}^{n},\ldots,Y_{K,(1,P)}^{n}|Y_{1,P}^{n},\Omega\right)\nonumber\\
&\quad + h(Y_{2,(1,NP)}^{n}, \ldots, Y_{K,(1,NP)}^{n}| Y_{1}^{n},\tilde{Y}_{1}^{n},\ldots, \tilde{Y}_{M-1}^{n} ,W_{1},\Omega)\\
&\leq h(Y_{1,P}^{n}|W_{1},\Omega) +  h(Y_{1,NP}^{n},\tilde{Y}_{1,NP}^{n},\ldots, \tilde{Y}_{M-1,NP}^{n} |W_{1},Y_{1,P}^{n}, \Omega) +n\epsilon_{n}\nonumber\\
&\quad + h\left(Y_{2,(1,P)}^{n},\ldots,Y_{K,(1,P)}^{n}\right|Y_{1,P}^{n},\Omega)\nonumber\\
&\quad + \underbrace{h(Y_{2,(1,NP)}^{n}, \ldots, Y_{K,(1,NP)}^{n}| Y_{1,NP}^{n},\tilde{Y}_{1,NP}^{n},\ldots, \tilde{Y}_{M-1,NP}^{n} ,\Omega)}_{\leq no(\log(P))}\label{Eq:use}.
\end{align}

In (\ref{Eq:use}), we used the fact that given $\left(Y_{1,NP}^{n},\tilde{Y}_{1,NP}^{n},\ldots, \tilde{Y}_{M-1,NP}^{n}\right)$ and $\Omega$, the channel input $X^{n}_{1,NP}$ can be obtained within noise distortion and hence the channel outputs $\left(Y^{n}_{2,(1,NP)},\ldots, Y^{n}_{K,(1,NP)}\right)$ can be obtained within noise distortion.

Adding (\ref{Eq:R1bound}) and (\ref{Eq:use}), we obtain
\begin{align}
&n\Big[MR_{1}+R_{2}+\cdots+R_{K}\Big]\nonumber\\
&\leq nM\log(P) + h\left(Y_{2,(1,P)}^{n},\ldots,Y_{K,(1,P)}^{n}\right|Y_{1,P}^{n},\Omega) + n\epsilon^{'}_{n}+no(\log(P)).\label{EAmain}
\end{align}

We next note that each of the $Y_{2,(1,P)}^{n}, \ldots, Y_{K,(1,P)}^{n}$ sequences are of length $\lambda n$. We proceed to upper bound the second term in (\ref{EAmain}) by considering two cases:

If $\min(M,K)=K$, then we have the following bound:
\begin{align}
h\left(Y_{2,(1,P)}^{n},\ldots,Y_{K,(1,P)}^{n}\right|Y_{1,P}^{n}, \Omega) &\leq \sum_{i=1}^{\lambda n} h(Y_{2,(1,P)}(i), \ldots, Y_{K,(1,P)}(i))\\
&\leq \sum_{i=1}^{\lambda n} \sum_{j=2}^{K}h(Y_{j,(1,P)}(i))\\
&\leq n(K-1)\lambda \log(P)\label{EA1}
\end{align}

If $\min(M,K)=M$, then we proceed as follows:
\begin{align}
&h\left(Y_{2,(1,P)}^{n},\ldots,Y_{K,(1,P)}^{n}\right|Y_{1,P}^{n}, \Omega) \nonumber\\
&\leq \sum_{i=1}^{\lambda n} h(Y_{2,(1,P)}(i), \ldots, Y_{K,(1,P)}(i)|Y_{1,P}(i),\Omega)\\
&= \sum_{i=1}^{\lambda n} h(Y_{2,(1,P)}(i),\ldots, Y_{M, (1,P)}(i)|Y_{1,P}(i),\Omega) \nonumber\\
&\quad + \sum_{i=1}^{\lambda n}h(Y_{M+1,(1,P)}(i),\ldots, Y_{K,(1,P)}(i)|Y_{1,(1,P)}(i),\ldots, Y_{M,(1,P)}(i),\Omega)\\
&\leq n(M-1)\lambda \log(P)\nonumber\\
&\quad + \sum_{i=1}^{\lambda n}\underbrace{h(Y_{M+1,(1,P)}(i),\ldots, Y_{K,(1,P)}(i)|Y_{1,(1,P)}(i),\ldots, Y_{M,(1,P)}(i),\Omega)}_{\leq o(\log(P))}\label{Eq:last}\\
&\leq n(M-1)\lambda \log(P)+n\lambda o(\log(P)).\label{EA2}
\end{align}
where in (\ref{Eq:last}), we used the fact that given $(Y_{1,(1,P)}(i),\ldots, Y_{M,(1,P)}(i),\Omega)$, the channel input $X_{1,P}(i)$ can be obtained within noise distortion via channel inversion. 
To note this, we can write
\begin{align}
\left[
\begin{array}{c}
Y_{1,P}(i)\\
Y_{2, (1,P)}(i)\\
\vdots\\
Y_{M,(1,P)}(i)
\end{array}
\right]&= 
\underbrace{\left[
\begin{array}{c c c}
H^{1}_{1, P}(i) \quad\cdots \quad H^{M}_{1,P}(i) \\
H^{1}_{2, (1,P)}(i) \quad\cdots \quad H^{M}_{2,(1,P)}(i) \\
\vdots   \quad \ddots  \quad \vdots \\
H^{1}_{M, (1,P)}(i) \quad\cdots \quad H^{M}_{M,(1,P)}(i) \\
\end{array}
\right]}_{\text{Full rank}}\left[
\begin{array}{c}
X^{1}_{1,P}(i)\\
X^{2}_{1,P}(i)\\
\vdots\\
X^{M}_{1,P}(i)
\end{array}
\right]+ 
\left[
\begin{array}{c}
N_{1,P}(i)\\
N_{2,(1,P)}(i)\\
\vdots\\
N_{M,(1,P)}(i)
\end{array}
\right].
\end{align}
Thus, the channel input  $X_{1,P}(i)= [X^{1}_{1,P}(i), \ldots, X^{M}_{1,P}(i)]^{T}$ at time $i$ can be obtained via channel inversion. 
Subsequently $(Y_{M+1,(1,P)}(i),\ldots, Y_{K,(1,P)}(i))$ can obtained within noise distortion from $X_{1,P}(i)$ and $\Omega$. 
Therefore, from (\ref{EA1}) and (\ref{EA2}), we conclude that 
\begin{align}
h\left(Y_{2,(1,P)}^{n},\ldots,Y_{K,(1,P)}^{n}\right|Y_{1,P}^{n}, \Omega) &\leq n(\min(M,K)-1)\lambda \log(P) + n\lambda o(\log(P)),\label{EA}
\end{align}
Hence, from (\ref{EAmain}) and (\ref{EA}), we have  
\begin{align}
MR_{1}+R_{2}+\ldots + R_{K}&\leq [M+ (\min(M,K)-1)\lambda]\log(P) + \epsilon^{''}_{n} + o(\log(P))
\end{align}
Normalizing by $\log(P)$ and taking the limits $n\rightarrow \infty$ and then $P\rightarrow \infty$, we obtain 
\begin{align}
Md_{1}+d_{2}+\ldots + d_{K}&\leq M+ (\min(M,K)-1)\lambda.
\end{align}

\begin{remark}
We note here that the bound 
\begin{align}
Md_{1}+d_{2}+\ldots + d_{K}&\leq M+ (\min(M,K)-1)\lambda
\end{align}
can be tightened to 
\begin{align}
\min(M,K)d_{1}+d_{2}+\ldots + d_{K}&\leq \min(M,K)+ (\min(M,K)-1)\lambda\label{improved}
\end{align}
under additional assumptions on the system model. In particular, recall that in the proof, we had created $(M-1)$ artificial receivers (that were statistically equivalent to receiver $1$). However, if the channel coherence patterns of all the users are identical, then one can restrict the number of transmit antennas to $\min(M,K)$. Subsequently, the converse proof can be modified by creating $(\min(M,K)-1)$ number of artificial receivers and resulting in an improved bound of (\ref{improved}). However, it is worth mentioning that the main result of Theorem \ref{Theorem1}, i.e., the minimum fraction of perfect CSIT required to achieve maximum DoF would remain the same under the strengthening of the bound. 
\end{remark}

\bibliographystyle{unsrt}
\bibliography{refravi}

\begin{thebibliography}{10}

\bibitem{MIMOBC}
H.~Weingarten, Y.~Steinberg, and S.~Shamai.
\newblock The capacity region of the {G}aussian multiple-input multiple-output
  broadcast channel.
\newblock {\em IEEE Trans. Inf. Theory}, 52(9):3936--3964, Sept. 2006.

\bibitem{CaireShamai}
G.~Caire and S.~Shamai.
\newblock On the achievable throughput of a multiantenna {G}aussian broadcast
  channel.
\newblock {\em IEEE Trans. Inf. Theory}, 49(7):1691--1706, July 2003.

\bibitem{NoCSITJafar}
C.~Huang, S.~A. Jafar, S.~Shamai, and S.~Viswanath.
\newblock On degrees of freedom region of {M}{I}{M}{O} networks without channel
  state information at transmitters.
\newblock {\em IEEE Trans. Inf. Theory}, 58(2):849--857, Feb. 2012.

\bibitem{VV:NOCSIT}
C.~S. Vaze and M.~K. Varanasi.
\newblock The degrees of freedom regions of {M}{I}{M}{O} broadcast,
  interference, and cognitive radio channels with no {C}{S}{I}{T}, [ar{X}iv:
  0909.5424].
\newblock {\em \emph{Submitted to} IEEE Trans. Inf. Theory}, Sept. 2009.

\bibitem{JafarGoldsmithIsotropic}
S.~A. Jafar and A.~Goldsmith.
\newblock Isotropic fading vector broadcast channels: The scalar upperbound and
  loss of degrees of freedom.
\newblock {\em IEEE Trans. Inf. Theory}, 51(3):848--857, March 2005.

\bibitem{MaddahAli-Tse:DCSI-BC}
M.~A. Maddah-Ali and D.~Tse.
\newblock Completely stale transmitter channel state information is still very
  useful.
\newblock {\em IEEE Trans. Inf. Theory}, 58(7):4418--4431, July 2012.

\bibitem{VV:DCSI-BC}
C.~S. Vaze and M.~K. Varanasi.
\newblock The degrees of freedom regions of two-user and certain three-user
  {M}{I}{M}{O} broadcast channels with delayed {C}{S}{I}{T} [ar{X}iv:
  1101.0306v2].
\newblock {\em \emph{Submitted to} IEEE Trans. Inf. Theory}, Dec. 2011.

\bibitem{Weingarten_Shamai_Kramer}
H.~Weingarten, S.~Shamai, and G.~Kramer.
\newblock On the compound {MIMO} broadcast channel.
\newblock In {\em Proceedings of Information Theory and Applications Workshop
  UCSD}, La Jolla, CA, Jan. 2007.

\bibitem{CompoundJafar}
T.~Gou, S.~A. Jafar, and C.~Wang.
\newblock On the degrees of freedom of finite state compound wireless networks.
\newblock {\em IEEE Trans. Inf. Theory}, 57(6):3286--3308, June 2011.

\bibitem{MAMA:CompoundBC}
M.~A. Maddah-Ali.
\newblock On the degrees of freedom of the compound {M}{I}{M}{O} broadcast
  channels with finite states, [ar{X}iv: 0909.5006v3].
\newblock {\em \emph{Submitted to} IEEE Trans. Inf. Theory}, Oct. 2009.

\bibitem{Jindal_BCFB}
N.~Jindal.
\newblock {MIMO} broadcast channels with finite rate feedback.
\newblock {\em IEEE Trans. Inf. Theory}, 51(5):5045--5049, Nov. 2006.

\bibitem{Caire_Jindal_Shamai}
G.~Caire, N.~Jindal, and S.~Shamai.
\newblock On the required accuracy of transmitter channel state information in
  multiple antenna broadcast channels.
\newblock In {\em Proceedings of the Asilomar Conference on Signals, Systems
  and Computers}, Pacific Grove, CA, 2007.

\bibitem{Kobayashi_Caire_Jindal}
M.~Kobayashi, G.Caire, and N.~Jindal.
\newblock How much training and feedback are needed in {MIMO} broadcast
  channels?
\newblock In {\em Proceedings of the IEEE International Symposium on
  Information Theory}, pages 2663--2667, Toronto, Canada, Aug. 2008.

\bibitem{JafarTCBC}
T.~Gou and S.~A. Jafar.
\newblock Optimal use of current and outdated channel state information-
  degrees of freedom of the {M}{I}{S}{O} {B}{C} with mixed {C}{S}{I}{T}.
\newblock {\em IEEE Communications Letters}, 16(7):1084--1087, July 2012.

\bibitem{KobayashiTCBC}
S.~Yang, M.~Kobayashi, D.~Gesbert, and X.~Yi.
\newblock Degrees of freedom of time correlated {M}{I}{S}{O} broadcast channel
  with delayed {C}{S}{I}{T}, [ar{X}iv: 1203.2550].
\newblock {\em \emph{Submitted to} IEEE Trans. Inf. Theory}, Mar. 2012.

\bibitem{EliaMixed}
J.~Chen and P.~Elia.
\newblock Degrees-of-freedom region of the {M}{I}{S}{O} broadcast channel with
  general mixed-{C}{S}{I}{T}, [ar{X}iv: 1205.3474].
\newblock May 2012.

\bibitem{Jafar_corr}
Syed~A. Jafar.
\newblock {Blind interference alignment}.
\newblock {\em IEEE Journal of Selected Topics in Signal Processing},
  6(3):216--227, June 2012.

\bibitem{RetroIA}
H.~Maleki, S.~A. Jafar, and S.~Shamai.
\newblock Retrospective interference alignment over interference networks.
\newblock {\em IEEE Journal of Selected Topics in Signal Processing},
  6(3):228--240, June 2012.

\bibitem{Wang_Gou_Jafar}
Tiangao Gou, Chenwei Wang, and Syed~A. Jafar.
\newblock Aiming perfectly in the dark - blind interference alignment through
  staggered antenna switching.
\newblock {\em IEEE Trans. Signal Processing}, 59:2734--2744, June 2011.

\bibitem{TandonJafarShamaiPoor:alternating}
R.~Tandon, Syed~A. Jafar, S.~Shamai, and H.~V. Poor.
\newblock On the synergistic benefits of alternating {C}{S}{I}{T} for the
  {M}{I}{S}{O} {B}{C}.
\newblock {\em IEEE Trans. Inf. Theory, submitted, ar{X}iv:1208.5071}, Aug.
  2012.

\bibitem{NamyoonHeath}
N.~Lee and R.~W.~Heath Jr.
\newblock Not too delayed {C}{S}{I}{T} achieves the optimal degrees of freedom,
  [ar{X}iv: 1207.2211].
\newblock July 2012.

\bibitem{ElGamalFB}
A.~El Gamal.
\newblock The feedback capacity of degraded broadcast channels.
\newblock {\em IEEE Trans. Inf. Theory}, 24(3):379--381, May 1978.

\end{thebibliography}
\end{document}